\documentclass[preprint]{revtex4}
\usepackage{graphicx}
\usepackage{epsfig}
\usepackage{amsmath}
\usepackage{float}
\usepackage{subfigure}
\usepackage{wrapfig}
\usepackage{multirow}
\usepackage{tabularx}
\usepackage{array}
\usepackage{url}
\usepackage{amssymb}
\usepackage{systeme}
\usepackage{braket}
\usepackage{caption}
\usepackage{dcolumn}
\usepackage{bm}
\usepackage{amsfonts}
\usepackage{xcolor}

\begin{document}

\title{Tidal Deformability of Quark Stars with Repulsive Interactions}

\author{M. B. Albino\dag\,, R. Fariello\dag\,\ddag\, and F. S. Navarra\dag\ }
\address{\dag\ Instituto de F\'{\i}sica, Universidade de S\~{a}o Paulo\\
Rua do Mat\~ao, 1371, 05508-090, Cidade Universit\'aria, S\~{a}o Paulo, SP, Brazil} 
\address{\ddag\ Departamento de Ci\^encias da Computa\c{c}\~ao,
  Universidade Estadual de Montes Claros, Avenida Rui Braga, sn,
  Vila Mauric\'eia, Montes Claros, 39401-089, Minas Gerais, Brazil}

\begin{abstract}
In an early work, we applied a QCD-based EOS to
the study of the stellar structure of self-bound strange stars, obtaining
sequences with maximum masses larger than two solar masses and radii
ranging from 8 to 12 Km. In this work, we update the previous
calculations and compare them with the most recent data, including the very 
recent determination of the mass and radius of the massive pulsar PSR  
J0740+6620 performed by the NICER and XMM-Newton Collaborations. 
Our equation of state is similar to the MIT bag model one,
but it includes repulsive interactions, which turn out to be essential to
reproduce the accumulated experimental information. We find that our EOS is 
still compatible with all astrophysical observations but the parameter window 
is now narrower.

\end{abstract}

\pacs{PACS Numbers : 97.10.Cv, 12.38.-t, 12.38.Mh }
\maketitle



\vspace{1cm}
\section{Introduction}

The region of the QCD phase diagram 
with low temperature and high chemical potential is still not well understood.
From the theoretical point of view, there are no accurate first principles
predictions
for the properties of QCD matter at high baryon densities. The numerical
lattice
simulation techniques that have been successfully applied to the study of the hot
quark gluon plasma (QGP) fail in the cold, baryon rich conditions due to the
sign problem. In spite  
of the difficulties, significant progress has been accomplished both in the
theoretical description of moderate density nuclear matter \cite{eosnuc,tkhs}
and ultrahigh-density matter
\cite{gorda21,vuku20} but no reliable results exist in the crucial regime
between
approximately one and ten nuclear saturation densities. Several model
calculations
suggest that there is a low temperature deconfined phase of quarks and gluons,
the cold QGP, also called quark matter (QM). This phase might exist in the 
the core of dense stars, an idea that has been around already for some
decades \cite{decs,alf}. It is even possible that a whole star, not only its
core, be made of quark matter \cite{witten84}. This possibility was
explored in several works
\cite{qstars,jorge20,fran12,jorge11,pagliara2011} and will be further
explored in this work, which is an update of \cite{fran12}. 

An early analysis of the existing observational data 
presented in \cite{ozel2006} concluded that most of the QM
EOSs were too soft and therefore unable to support the existence of
neutron stars with a quark phase. Since then it was shown in several
works that a self-bound star, composed entirely of quark matter, could
explain a massive neutron star. In order to obtain a stiff enough quark matter
equation of state, several groups introduced repulsive interactions among
the quarks, mediated by the exchange of vector particles
\cite{fran12,vbag,baym19,deb20,oert20,pisa21,sylhz}, which can be
``effective massive gluons'' or ``effective vector mesons''. Interestingly,
most of these developments make use of a mean field approximation for the
vector field and arrive at a similar result, which is a
quadratic term in the baryon density present both in the pressure and energy density. 


From the experimental side, during the last decade we have witnessed
remarkable advances in the
observation of neutron stars: 
the discovery of extremely massive neutron stars \cite{demo10,antoniadis};  
qualitative improvements in X-ray radius measurements
\cite{gui13,oz16,freire16,stein17,stein16,miller17,bogda16};  
and the famous LIGO/Virgo detection of gravitational waves (GWs)
originating from the NS-NS merger GW170817 \cite{ligo17}.  
Increasingly stringent constraints have been placed on the EOS of
NS matter.  
An accurate measurement of a compact object using Shapiro
delay \cite{croma}
yielded $ 2.14^{+0.1}_{-0.09} \,\,  M_{\odot}$ for the J0740+6620 pulsar.
It has been argued that a handful of compact stars may achieve masses greater
than the PSR J0740+6620. 
The events denoted as GW190814 \cite{ligo20} and GW190425 \cite{ligo20a}   
suggest that the NS mass can be larger than
$ 2.5  \,\,  M_{\odot}$. 
Recent data (including a reliable determination of the radius) about the 
pulsar PSR J0030+0451 were published in Refs. 
\cite{nicer1,nicer2,nicer3,nicer4,nicer5}. Finally, very recently \cite{nicer6} 
the NICER and XMM-Newton Collaborations presented a determination of the 
mass and radius of the massive pulsar PSR J0740+6620.

Differences between candidate EOSs can have a significant
effect on the tidal interactions of neutron stars. 
Recently new constraints appeared on the tidal deformability \cite{ligo18}. 
It has been realized \cite{annala} that         
the two-solar-mass constraint forces the EOS to be relatively
stiff at low densities. At the same time, the constraint on
$\Lambda(1.4 \, M_{\odot})$ sets an upper limit for the stiffness,
constraining the EOS band in a complementary direction.


In this paper we will update the study presented in \cite{fran12} and check
whether the EOS introduced in \cite{davi} remains a viable option, satisfying
the most recent experimental contraints.

This text is organized as follows. In Sec. II we briefly review the EOS for
the cold QGP. In Sec. III we introduce the
stability conditions and discuss its consequences. In Sec. IV we present the
Tolman-Oppenheimer-Volkoff (TOV) equations for stellar
structure calculations and their numerical solutions. In Sec. V we discuss the
tidal deformability and in Sec. VI we present some comments and conclusions.

\section{The equation of state}

Following \cite{fran12}, 
we consider a quark star consisting of {\it u}, {\it d} and {\it s} quarks
with masses $m_{u}=$ $5$ MeV, $m_{d}=$ $7$ MeV, and $m_{s}=$ $100$ MeV. 
The derivation of the EOS \cite{davi} used here starts with the assumption
that the gluon field can be
decomposed into low (``soft'') and high
(``hard'') momentum components. The expectation values of the soft fields were
identified with the gluon condensates of dimension two and
four, respectively. The former generates a dynamical mass, $m_G$ for the hard
gluons,
and the latter yields an analogue of the ``bag constant'' term
in the energy density and pressure.  Given the large number of quark sources,
even in
the weak coupling regime, the hard gluon fields
are strong, the occupation numbers are large, and therefore these fields can be
approximated by classical color fields.
The effect of the condensates is to soften the EOS whereas the hard gluons
significantly
stiffen it, by increasing both the energy density and pressure. With these
approximations it was possible to derive \cite{davi} an analytical expression
for the EOS, called MFTQCD (Mean Field Theory of QCD). When adapting this 
equation of state to the stellar medium, we assume, as usual, that quarks and 
electrons are in chemical equilibrium maintained by the weak 
processes \cite{farhi}.
Neutrinos are assumed to escape and do not contribute to the pressure and
energy density. 
Moreover, we impose charge neutrality and baryon number conservation.
These requirements yield a set of four algebraic equations for Fermi 
momentum calculation for each quark flavor ($u$, $d$ and $s$) and for the 
electrons ($e$)
$$
{k_{u}}^{3}+{k_{d}}^{3}+{k_{s}}^{3}=3\pi^{2}\rho_{B},
$$
\begin{equation}
2{k_{u}}^{3}={k_{d}}^{3}+{k_{s}}^{3}+{k_{e}}^{3},
\label{densidadee}
\end{equation}
$$
{k_{d}}^{2}+{m_{d}}^{2}={k_{s}}^{2}+{m_{s}}^{2},
$$
$$
\sqrt{{k_{u}}^{2}+{m_{u}}^{2}}+\sqrt{{k_{e}}^{2}+{m_{e}}^{2}}
=\sqrt{{k_{s}}^{2}+{m_{s}}^{2}},
$$
for a fixed baryon density $\rho_{B}$. The energy density is given
by \cite{davi}
$$
\epsilon=\bigg({\frac{27g^{2}}{2{m_{G}}^{2}}}\bigg) \ {\rho_{B}}^{2} +
\mathcal{B}_{QCD}
$$
$$
+\sum_{i=u,d,s}3{\frac{\gamma_{Q}}{2{\pi}^{2}}} \Bigg\lbrace
{\frac{{k_{i}}^{3}\sqrt{{k_{i}}^{2}+{m_{i}}^{2}}}{4}} +
{\frac{{m_{i}}^{2}{k_{i}}\sqrt{{k_{i}}^{2}+{m_{i}}^{2}}}{8}} -
{\frac{{m_{i}}^{4}}{8}}ln\Big[{k_{i}}+\sqrt{{k_{i}}^{2}+{m_{i}}^{2}} \ \Big]
+ {\frac{{m_{i}}^{4}}{16}}ln({m_{i}}^{2}) \Bigg\rbrace
$$
\begin{equation}
  +{\frac{\gamma_{e}}{2{\pi}^{2}}} \Bigg\lbrace
  {\frac{{k_{e}}^{3}\sqrt{{k_{e}}^{2}+{m_{e}}^{2}}}{4}} +
  {\frac{{m_{e}}^{2}{k_{e}}\sqrt{{k_{e}}^{2}+{m_{e}}^{2}}}{8}}
  - {\frac{{m_{e}}^{4}}{8}}ln\Big[{k_{i}}+\sqrt{{k_{e}}^{2}+{m_{e}}^{2}}
    \ \Big] + {\frac{{m_{e}}^{4}}{16}}ln({m_{e}}^{2}) \Bigg\rbrace,
\label{epsib}
\end{equation}
and the pressure is
$$
p=\bigg({\frac{27g^{2}}{2{m_{G}}^{2}}}\bigg) \ {\rho_{B}}^{2}
-   \mathcal{B}_{QCD}
$$
$$
+\sum_{i=u,d,s}{\frac{\gamma_{Q}}{2{\pi}^{2}}} \Bigg\lbrace
{\frac{{k_{i}}^{3}\sqrt{{k_{i}}^{2}+{m_{i}}^{2}}}{4}}  -
{\frac{3{m_{i}}^{2}{k_{i}}\sqrt{{k_{i}}^{2}+{m_{i}}^{2}}}{8}}
+ {\frac{3{m_{i}}^{4}}{8}}ln\Big[
  {k_{i}}+\sqrt{{k_{i}}^{2}+{m_{i}}^{2}} \ \Big]-
{\frac{3{m_{i}}^{4}}{16}}ln({m_{i}}^{2}) \Bigg\rbrace
$$
\begin{equation}
  +{\frac{\gamma_{e}}{6{\pi}^{2}}} \Bigg\lbrace
  {\frac{{k_{e}}^{3}\sqrt{{k_{e}}^{2}+{m_{e}}^{2}}}{4}}
  - {\frac{3{m_{e}}^{2}{k_{e}}\sqrt{{k_{e}}^{2}+{m_{e}}^{2}}}{8}}
  + {\frac{3{m_{e}}^{4}}{8}}ln\Big[
    {k_{e}}+\sqrt{{k_{e}}^{2}+{m_{e}}^{2}} \ \Big] -
  {\frac{3{m_{e}}^{4}}{16}}ln({m_{e}}^{2}) \Bigg\rbrace ,
\label{pressb}
\end{equation}
where $m_{e}=$ $0.5$ MeV is the electron mass, $m_{G}$ is the dynamical
gluon mass, and $g$ is the coupling constant $(\alpha_{s}=g^{2}/4\pi)$ in QCD.
Our analogue of the bag constant, called here $\mathcal{B}_{QCD}$, is given by
\begin{equation}
  \mathcal{B}_{QCD}= \frac{9}{128} \, \phi_{0}^4 =
  \langle \frac{1}{4} F^{a\mu\nu}F^{a}_{\mu\nu} \rangle,
\label{bag}
\end{equation}
where $\phi_0$ is an energy scale associated with the energy density of the
vacuum and with the gluon condensate \cite{davi}.
In (\ref{epsib}) and (\ref{pressb}) the summation over quark colors has
already been performed. Throughout this work we employ the natural
units $G=1$, $\hbar=1$, $c=1$. Comparing Eqs. (\ref{epsib}) and (\ref{pressb}) 
with the equivalent definitions of energy and pressure in the modified
bag model with postulated repulsive vector interactions (see Eqs. (14)
and (16) of Ref. \cite{bla20}), we observe a similarity. Both
EOSs have a term proportional to $\rho_B^2$. In \cite{davi} it was derived
from QCD whereas in \cite{bla20} it was postulated.

\section{Stability conditions}

In this section we discuss the two stability conditions, which have to be
satisfied by stable strange quark matter. The first one is that the energy
per baryon of the deconfined phase
(for $P=0$ and $T=0$) is lower than the nonstrange infinite baryonic
matter defined in \cite{farhi,pagliara2011}. Following these works we
impose that:
\begin{equation}
E_{A} \equiv \frac{\epsilon}{\rho_{B}} \leq  934  \,\,\,  \mbox{MeV}.
\label{estabilidade}
\end{equation}
This condition must hold at the zero pressure point and hence we can, from
(\ref{epsib}) and (\ref{pressb}), numerically
derive a relation between the bag constant $B_{QCD}$ and the ratio
$\xi=g/m_{G}$. We solve (\ref{pressb}) obtaining
$\rho_B=\rho_B (B_{QCD}, \xi)$, which is then inserted into (\ref{epsib}).
The resulting expression is used to write the
condition $\epsilon (B_{QCD}, \xi)/ \rho_B (B_{QCD}, \xi) = 934$ MeV,
which defines one ``stability frontier''.
This last equation is rewritten as $\xi = \xi (B_{QCD})$, is plotted in
Fig. 1 (solid line) and denoted by the 3-flavor line.
Points in the
$(\mathcal{B}_{QCD},\xi)$ plane located on the right of the solid
line are discarded since they do not satisfy
(\ref{estabilidade}). The solid line, corresponding to the maximal value
of $E_{A} = 934$ MeV, determines the maximum value of
$\mathcal{B}_{QCD}$. The minimum value of
$\mathcal{B}_{QCD}$ is determined by the
second stability condition, which requires nonstrange
quark matter in the bulk to have an energy per baryon higher than the one
of nonstrange infinite baryonic matter. By imposing that
\begin{equation}
E_{A} \equiv \frac{\epsilon}{\rho_{B}}   \geq  934  \,\,\,  \mbox{MeV},
\label{estabilidade2}
\end{equation}
for a two flavor quark matter at ground state, we ensure that atomic
nuclei do not dissolve into their constituent
quarks. The constraint (\ref{estabilidade2}) defines the dashed line in
the $(\mathcal{B}_{QCD},\xi)$ plane, denoted by the
2-flavor line in Fig. 1. Points located on the left of this line are
excluded because they do not satisfy (\ref{estabilidade2}).
The region between the two lines in Fig. 1 defines our stability window.
\begin{figure}[h]
\begin{center}
\epsfig{file=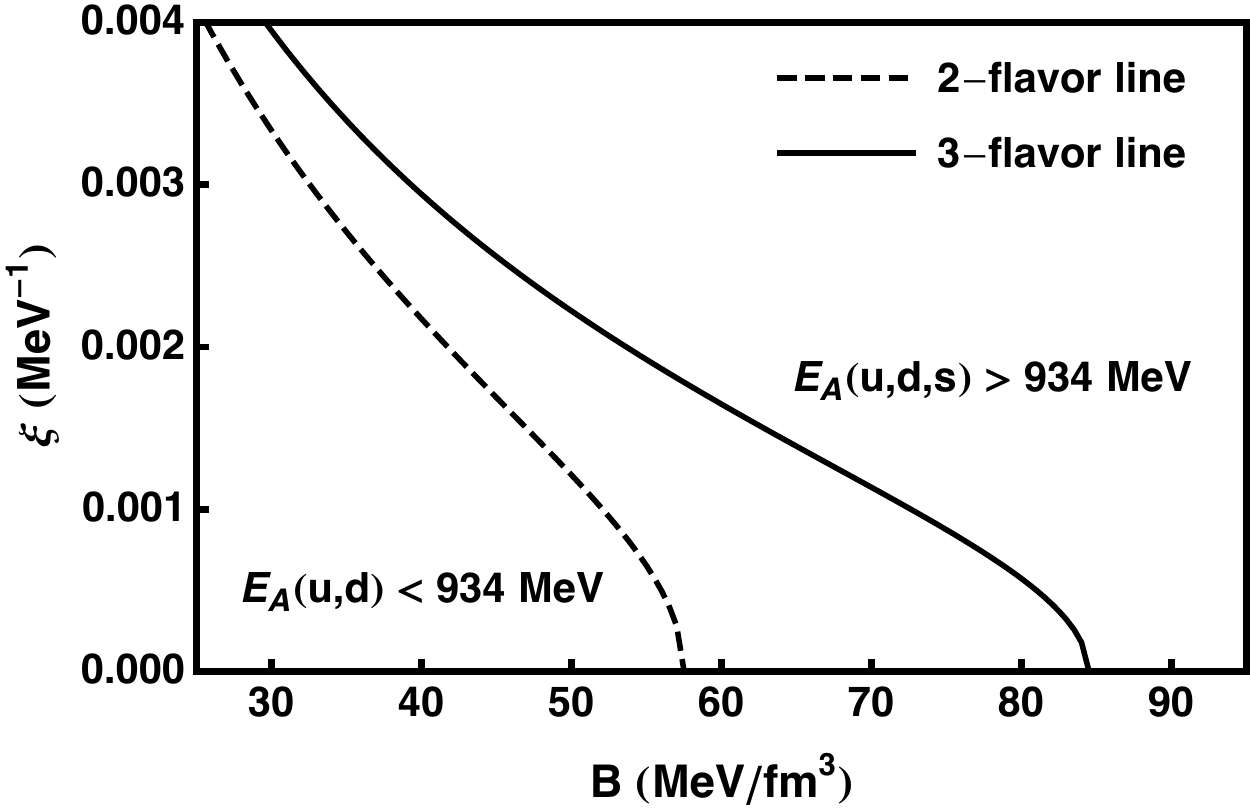,width=116mm}
\caption{Values of $\xi = g/m_{G}$ as a function of  $\mathcal{B}_{QCD}$
  for different values of the energy per baryon. The two lines define the
stability region.}
\end{center}
\label{fig1}
\end{figure}
Having fixed the $\mathcal{B_{QCD}}$ and $\xi$ parameters, we go back    
to (\ref{epsib}) and (\ref{pressb}) and, obtaining $\epsilon$ and $p$
for successive values of $\rho_B$,
we construct the EOS in the form $p = p(\epsilon)$, plotted in Fig. 2a.
In the figure, the different lines correspond to the three parameter sets 
listed in Table I. 
In this type of plot the slope is the speed of sound,
which, due to causality, can not exceed the unity. This limit is shown by the
full lines in the figure. In Fig. 2b we show the corresponding values of the 
speed of sound. As it can be seen, our model yields a much stiffer EOS, with 
a speed of sound much larger than the conformal value, for which 
$c^2_s = 1/3$. 
The dot-dashed line shows the EOS obtained from a recently  
updated version of the MIT bag model \cite{zzl,german}, which reads 
\begin{equation}
  p(\epsilon) = \frac{(\epsilon - B_{eff})}{3} -
  \frac{a_2^2}{12 \pi^2 a_4}
\left[1 + \sqrt{1 + \frac{16 \pi^2 a_4}{a_2^2} (\epsilon - B_{eff})} \right],
\label{mit20}
\end{equation} 
where $B_{eff}^{1/4} = 142.52 $ MeV,
$a_2^{1/2} = 100 $ MeV and $a_4 = 0.535$. As it can be seen, the MFTQCD EOS
generates stronger pressure for larger values of the parameter
$\xi = g/m_{G}$. This combination of parameters  
appears in the first term of (\ref{pressb}), which comes from the repulsive
interactions \cite{davi}.   
\begin{figure}[h]
\begin{tabular}{ccc}
 \includegraphics[width=.50\linewidth]{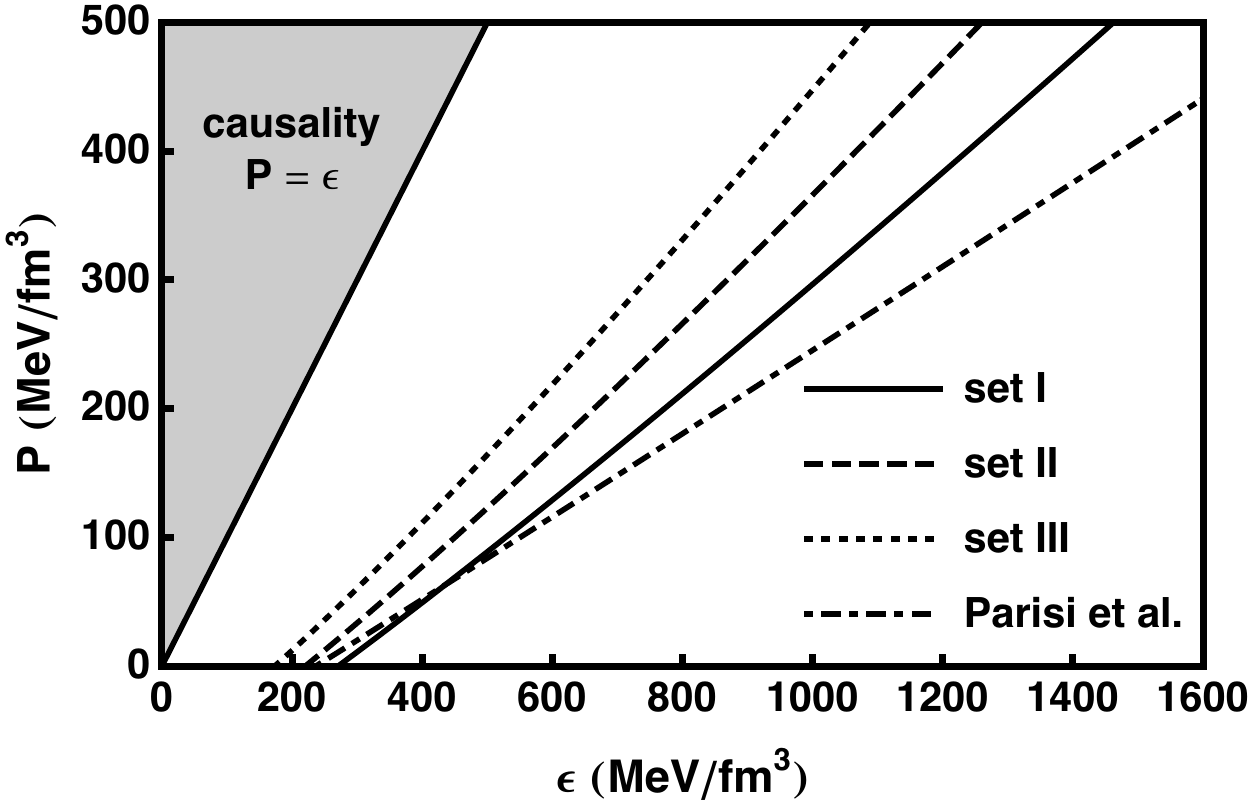}& \,\,\, &
 \includegraphics[width=.50\linewidth]{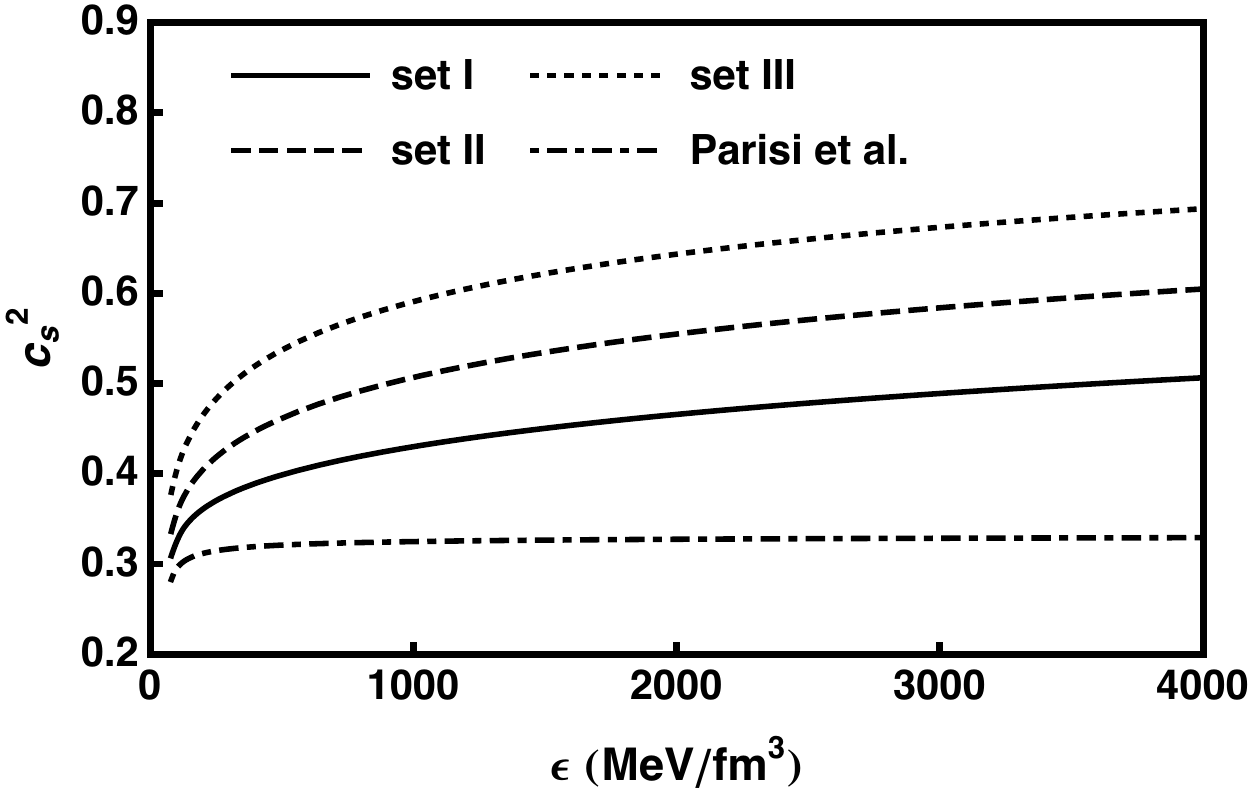} \\
  (a) & \,\,\, & (b)
\end{tabular}
\caption{a) Equation of state obtained with MFTQCD. Set I, II and III 
correspond to
the parameter combinations shown in Table I. 
For comparison,
  the dot-dashed line show the MIT Bag Model EOS used by Parisi et al.
  in Ref.~\cite{german}. b) Speed of sound for the same parameter choices.}
\label{fig2}
\end{figure}

\section{TOV equation, mass and radius}

In order to describe the structure of a static, non-rotating compact star, we
use the Tolman-Oppenheimer-Volkoff (TOV) equation for the pressure 
$p(r)$ \cite{glend}: 
\begin{equation}
  \frac{dp}{dr}=-\frac{\epsilon(r) M (r)}{r^2} \left[ 1 +
    \frac{p(r)}{\epsilon(r)} \right] \left[ 1 + \frac{4\pi r^3 p(r)}{M(r)}
    \right] \times
\left[ 1 - \frac{2M(r)}{r} \right]^{-1}.
\label{tov}
\end{equation}
The enclosed mass $M(r)$ of
the compact star is given by the mass continuity equation:
\begin{equation}
\frac{dM(r)}{dr}=4\pi r^2\epsilon(r).
\label{mass}
\end{equation}
Equations (\ref{tov}) and (\ref{mass}) express the balance between the
gravitational force and the internal pressure acting on a shell of mass
$dM(r)$ and thickness $dr$.

We solve numerically (\ref{tov}) and (\ref{mass}) for $p(r)$ and $M(r)$,
to obtain the mass-radius diagram.
The pressure and the energy density in (\ref{tov}) and (\ref{mass}) are given 
by the MFTQCD expressions (\ref{pressb}) and (\ref{epsib}), respectively.
We take the central energy density to be $\epsilon(r = 0)=\epsilon_{c}$ and
then we integrate out (\ref{tov}) and (\ref{mass}) from $r=0$ up to $r=R$,   
where the pressure at the surface is zero: $p(r=R)=0$. In Fig. 3 we show the 
mass-radius diagram for several values of $\mathcal{B_{QCD}}$ and $\xi$
respecting the stability condition.
\begin{figure}[h]
\vskip2mm 
\begin{center}
\epsfig{file=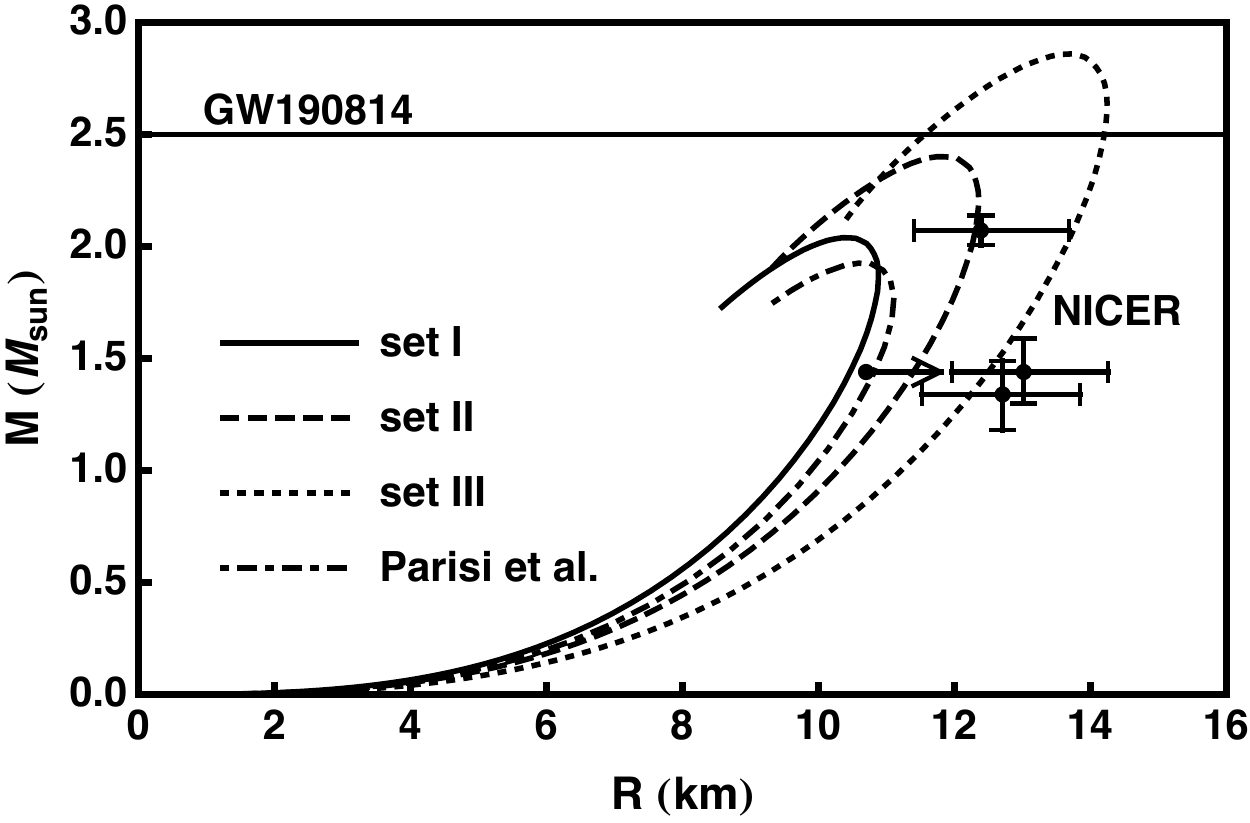,width=160mm}
\end{center}
\caption{Mass-radius diagram for combinations of  $\mathcal{B_{QCD}}$ and
$\xi$ allowed by the stability conditions. Set I, II and III correspond to 
the parameter combinations shown in Table I.  The points represent the region
favored by the measurements reported by the NICER and XMM-Newton 
Collaborations  
\cite{nicer1,nicer2,nicer3,nicer4,nicer5,nicer6}. The horizontal line 
shows the  mass of the compact object observed in the event GW190814.}  
\label{fig3}
\end{figure}
In the  diagram, the points represent the region
favored by the measurements reported in Refs.
\cite{nicer1,nicer2,nicer3,nicer4,nicer5,nicer6}. 
We can see that, with the parameters chosen in the indicated range, our EOS
is able to satisfy all the constraints shown in the mass-radius diagram.

\begin{table}[!htbp]
\caption{Parameter sets used in the figures.}
\vspace{0.3cm}
\centering
\begin{tabular}{ccc}
\hline
Set & $\mathcal{B_{QCD}} (MeV/fm^{3})$ & $\xi (MeV^{-1}) $
  \\ [0.8ex]
\hline
\hline
I   &  70    &  0.0011   \\
\hline
II  &  60    &  0.0016   \\
\hline
III &  50    &  0.0022    \\
\hline
\end{tabular}
\label{quarstar}
\end{table}

\section{Tidal deformability} 

An object that experiences the tidal force of another object will deform.
The susceptibility to deform is often measured using dimensionless quantities
that are called Love numbers. The Love number is an interesting
quantity because it can be used to probe the dense-matter EOS using data from
double-neutron-star-merger events. There are various binary systems where two 
objects orbit around each other. Because these objects loose energy to
gravitational waves, their orbits are not stable.
Therefore, they will inevitably approach each other until they finally merge. 
Around the collision point, the generated gravitational-wave signal is strong
enough to be detected by terrestrial instruments.

The tidal deformability parameter is given by \cite{tanja,tanja2,sabatucci} 
\begin{equation}
  \label{eq:tidal}
    \Lambda = \frac{2}{3} k_2 C^{-5},
\end{equation}
where $C \equiv M/R$ is the compactness of the star and $k_2$ is the tidal
Love number, which is given by \cite{tanja,tanja2,sabatucci}  
  \begin{eqnarray}
    k_{2} & = & \frac{8C^{5}}{5} \left(1-2C\right)^{2} \left[2+2C
              \left(y-1\right)-y\right] \left\{ 2C \left[6-3y+3C
              \left(5y-8\right)\right]\right. \nonumber\\
          & &\quad +4C^{3}\left[13-11y+C\left(3y-2\right)+2C^{2}
              \left(1+y\right)\right] \nonumber\\
          & &\quad \left.+3\left(1-2C\right)^{2}\left[2-y+2C
              \left(y-1\right)\right]
              \ln\left(1-2C\right)\right\}^{-1}, \label{eq:love}
  \end{eqnarray}
where
\begin{equation}\label{eq:y}
    y = \frac{R\, \beta(R)}{H(R)} - \frac{4\pi R^3 \epsilon_{sup}}{M},
          \quad \beta(r) = \frac{dH(r)}{dr}.
\end{equation}
In the above equation, the second term is a correction due to the fact that
in our model the energy density at the surface of the star,  
$\epsilon_{sup} \equiv \epsilon(P = 0)$ is not zero \cite{tanja2}. The functions
$H$ and $\beta$ can be obtained by solving the following system of differential
equations:
  \begin{eqnarray}
    H'(r) & = & \beta(r), \\
    \frac{d\beta}{dr} & = & 2
          \left(1 - \frac{2M(r)}{r} \right)^{-1} H(r)
          \left\{
            -2\pi
            \left[
              5\epsilon(r) +9P(r)
              +\frac{\epsilon(r)+P(r)}{dP/d\epsilon}
            \right]
          \right. \nonumber \\
          &&\left. \quad
            +\frac{3}{r^2} + 2\left(1 - \frac{2M(r)}{r} \right)^{-1}
            \left( \frac{M(r)}{r^2} +4\pi r P(r) \right)^2
          \right\} \nonumber \\
          &&\quad+\frac{2\beta(r)}{r}
          \left(1 - \frac{2M(r)}{r} \right)^{-1} 
          \left[
            -1 + \frac{M(r)}{r} +2\pi r^2
            \left( \epsilon(r) - P(r) \right)
          \right].\label{eq:beta}
  \end{eqnarray}

The Love number $k_2$ measures how easily the bulk
of the matter in a star is deformed. 
The Love number also encodes information about the star’s degree of central
condensation.
Stars that are more centrally condensed will have a smaller response to a tidal
field, resulting in a smaller Love number.
The Love number decreases with increasing compactness, and from
Eq.~(\ref{eq:love}) it can be seen that $k_2$ vanishes at the compactness 
of a black hole ($M/R = 0.5$) regardless of the EOS dependent quantity $y$.
The tidal Love numbers of strange quark matter stars
are qualitatively different from those of hadronic matter
stars \cite{tanja2,latt10,under20}. The latter decrease strongly for 
small values of the compactness.

In Fig.~4a we show the Love number $k_2$ as a function of the compactness 
$C$. 
We expect that a very compact star for any EoS is harder to deform then
a less compact one. This is what we see in the figures. 
It is interesting to observe that the same variation of $\mathcal{B_{QCD}}$
and $\xi$ which produces visible effects in the equation of state and in the
mass-radius diagram does not lead to appreciable differences in the
$k_2$-$C$ plot. The curves shown in Fig.~4a are practically identical 
to the curves 
in the analogous plots shown in Refs.~\cite{tanja2}, \cite{latt10} and
\cite{under20}, which were 
obtained with strange quark matter equations of state. This suggests that a wide
variety of quark matter EOSs lead to the same values of $k_2$. We also note 
that our curves are close to the one obtained with the
ultrarelativistic EOS with the speed of sound $c^2_s =1/3$ \cite{under20}. 
For completeness, in Fig.~4b we show the Love number $k_2$ as a function of 
the variable $y$.
\begin{figure}[h]                                                              
\begin{tabular}{ccc}                                                            
 \includegraphics[width=.50\linewidth]{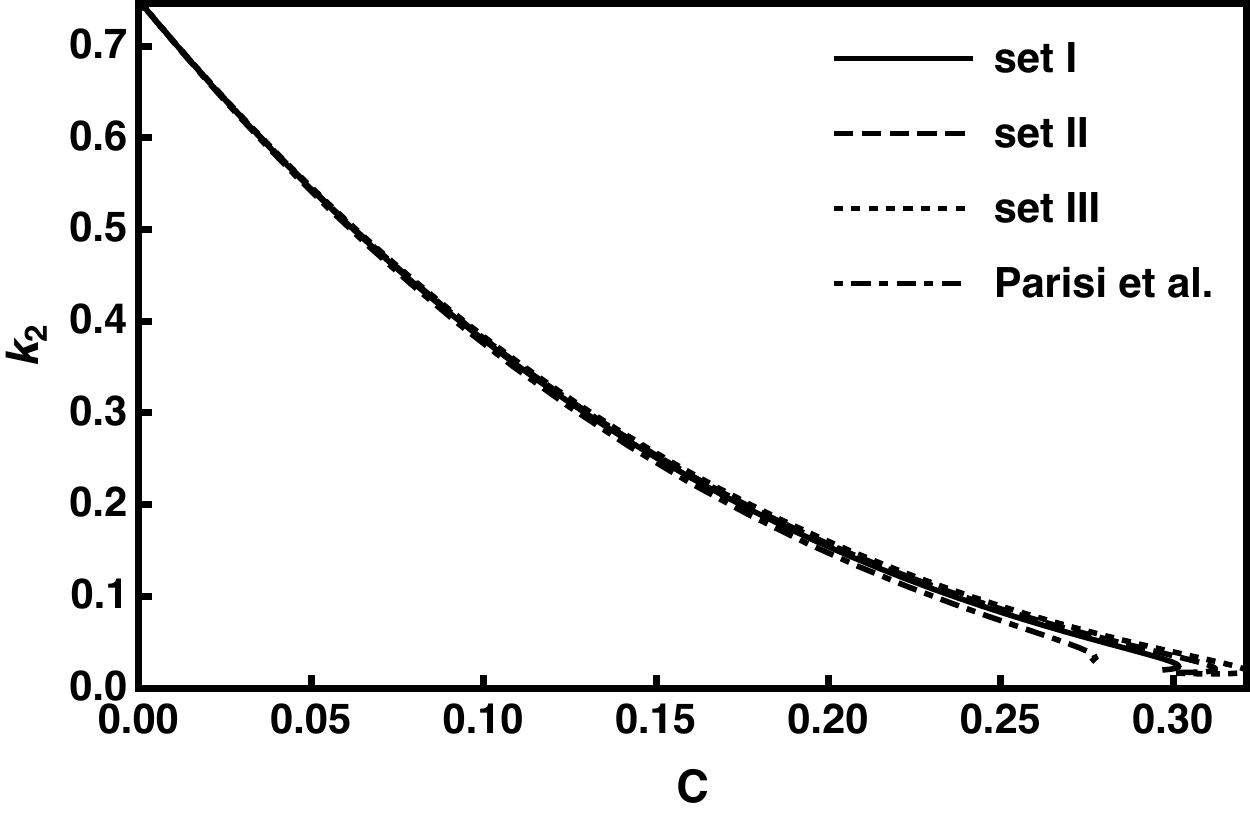}& \,\,\, &     
 \includegraphics[width=.50\linewidth]{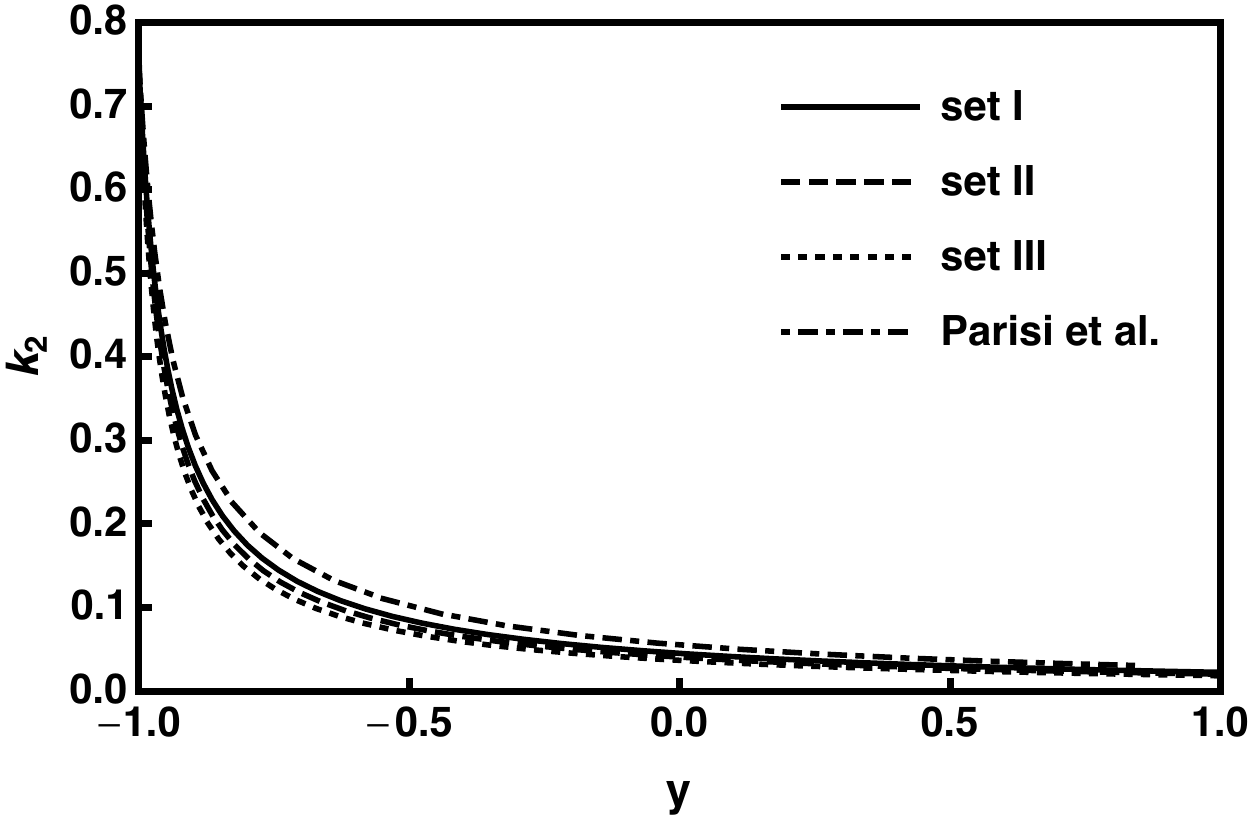} \\           
  (a) & \,\,\, & (b)                                            
\end{tabular}                                                     
\caption{a) Tidal Love number $k_2$ as a function of the compactness. b) $k_2$ 
as a function of $y$. The different lines correspond to the three 
parameter sets listed in Table I.}         
\label{fig4}                                                        
\end{figure}


As pointed out in \cite{tanja2}, in contrast to the Love number,
the tidal deformability has a wide range of values, spanning roughly
an order of magnitude over the observed mass range of neutron stars
in binary systems. The updated version of the tidal deformability
estimate for a $1.4 \, M_{\odot}$ 
neutron star based on the gravitational-wave event GW170817 \cite{ligo18}
implies that
\begin{equation}
  70 < \Lambda_{1.4} < 580.
  \label{lambdamax}
 \end{equation}
In Fig. 5 we show our results
for $\Lambda$ as a function of the star mass $M$.
As it can be seen, the constraint (\ref{lambdamax}) can be satisfied. 
We note, however, the visible tension between this constraint and those 
shown in the mass-radius plot. The larger values of the radius required 
to fit the NICER points seem to be somewhat difficult to reconcile with the 
$\Lambda$ values required by the GW170817 estimates. Other calculations 
performed with quark matter stars \cite{laura,zzl,wsz,zm,lz} or hybrid 
stars \cite{sylhz,consta,juwu,theo,kmt,han18} arrive at similar results. 
On the other hand, 
calculations of the tidal deformability with purely hadronic equations of  
states \cite{latt18,tan,latt21} seem to reproduce the experimental data 
more easily. 

\begin{figure}[h]
\begin{center}
\epsfig{file=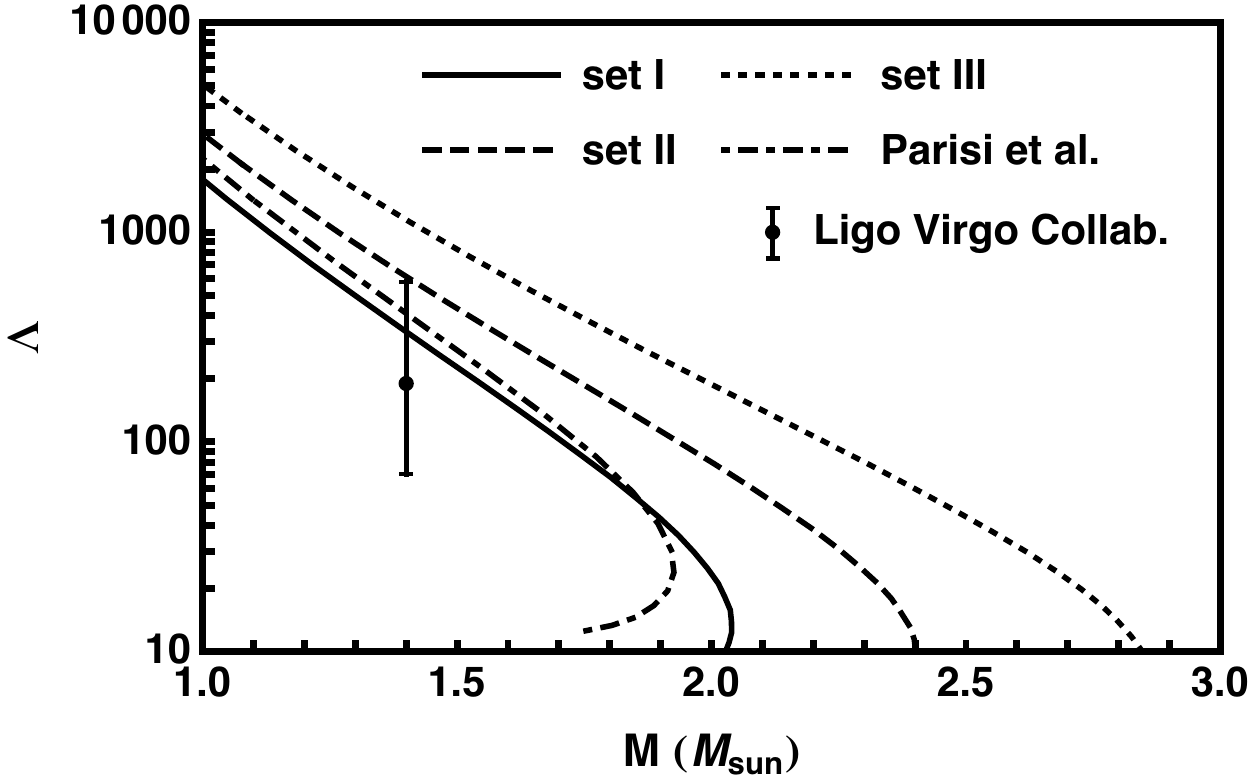,width=160mm}
\caption{The tidal deformability parameter $\Lambda$ as a function of the
  star mass. The different lines correspond to the three parameter sets
listed in Table I.
The vertical bar is the empirical tidal deformability at
  $M = 1.4 M_{\odot}$ inferred from the Bayesian analysis of the GW170817
  data at the 90\% confidence level \cite{ligo18}.}
\end{center}
\label{fig5}
\end{figure}

\section{Conclusion}

In \cite{davi} a new equation of state for cold quark matter was presented. 
It was soon applied to the the study of neutron stars, treated as 
self-bound strange quark stars. In this paper, almost ten years later,  
we have updated the calculations published in \cite{fran12} and checked 
whether that EOS can still account for the most recent astrophysical data. 
We find that MFTQCD is still a viable option. However, we observe that the  
parameter window is closing. A confirmation of the existing data and the 
reduction of the error bars in the tidal deformability and in the NICER 
neutron star radii data will be crucial to rule out strange quark star 
models and reduce the freedom in the choice of the equation of state.

\begin{acknowledgments}
We are deeply grateful to J.\ Horvath and to G.\ Lugones for fruitful discussions. 
This work was partially financed by the Brazilian funding agencies CAPES
and CNPq.
\end{acknowledgments}


\begin{thebibliography}{99}


\bibitem{eosnuc}  M. Oertel, M. Hempel, T.  Kl\"ahn
                  and S. Typel, Rev. Mod. Phys. {\bf 89}, 015007 (2017). 

\bibitem{tkhs}    I. Tews, T. Kr\"uger, K. Hebeler and A. Schwenk,
                  Phys. Rev. Lett. {\bf 110}, 032504 (2013).
  
\bibitem{gorda21} T.~Gorda, A.~Kurkela, R.~Paatelainen, S.~S\"appi
                  and A.~Vuorinen, arXiv:2103.07427 [hep-ph].
  
\bibitem{vuku20}  E.~Annala, T.~Gorda, A.~Kurkela, J.~N\"attil\"a and
                  A.~Vuorinen, 
                  Nature Phys. \textbf{16}, 907 (2020). 

\bibitem{decs}   D. Ivanenko and D. F. Kurdgelaidze,
                 Astrophysics J. {\bf 1}, 251 (1965);
                 D. Ivanenko and D. F. Kurdgelaidze,
                 Lett. Nuovo Cimento {\bf 2}, 13 (1969);
                 N. Itoh, Prog. Theor. Phys. {\bf 44}, 291 (1970).

\bibitem{alf}    M. Alford, D. Blaschke, A. Drago, T. Klahn,
                 G. Pagliara, and J. Schaffner-Bielich,
                 Nature {\bf  445}, E7 (2007).


                 
\bibitem{witten84}   E. Witten,  Phys. Rev. D {\bf 30}, 272  (1984).

\bibitem{qstars}   F. Weber, Prog. Part. and Nucl. Phys. {\bf 54}, 193 (2005);
                   R. X. Xu, S. I. Bastrukov, F. Weber, J. W. Yu,
                   and I. V. Molodtsova,
                   Phys. Rev. D {\bf 85}, 023008 (2012);
                   A. Drago, A. Lavagno, and I. Parenti,
                   Astrophys. J. {\bf 659}, 1519 (2007);
                   J. W. Yu and R. X. Xu,
                   Mon. Not. R. Astron. Soc. 414, 489 (2011);
                   A. Drago, A. Lavagno, and G. Pagliara,
                   Phys. Rev. D {\bf 89}, 043014 (2014);
                   A. Drago and G. Pagliara,
                   Phys. Rev. C {\bf 92}, 045801 (2015);
                   A. Drago, A. Lavagno, G. Pagliara, and D. Pigato,
                   Eur. Phys. J. A {\bf 52}, 40 (2016);
                   A. Drago and G. Pagliara, Eur. Phys. J. A {\bf 52}, 41 (2016);
                   G. Panotopoulos and I. Lopes,
                   Phys. Rev. D {\bf 96}, 083013 (2017);
                   G. Wiktorowicz, A. Drago, G. Pagliara, and S. B. Popov,
                   Astrophys. J. {\bf 846}, 163 (2017);
                   J. Bora and U. D. Goswami,  
                   Mon.\ Not.\ Roy.\ Astron.\ Soc.\  {\bf 502}, 1557 (2021).

\bibitem{jorge20}  J.~E.~Horvath and P.~H.~R.~S.~Moraes,
                   Int.\ J.\ Mod.\ Phys.\ D {\bf 30}, 2150016 (2021).
  
\bibitem{fran12}   B.~Franzon, D.~A.~Fogaca, F.~S.~Navarra and J.~E.~Horvath,
                   Phys.\ Rev.\ D {\bf 86}, 065031 (2012).

                   
\bibitem{jorge11}        M.G.B. de Avellar,  J. E. Horvath, and L. Paulucci,
                         Phys. Rev. D {\bf 84}, 043004  (2011).

\bibitem{pagliara2011}   S. Weissenborn, I. Sagert, G. Pagliara, M. Hempel,
                         and J. Schaffner - Bielich,
                         Astrophysics J. {\bf 740}, L14 (2011).
                         
\bibitem{ozel2006}       F. \"Ozel, Nature {\bf 441}, 1115 (2006).


\bibitem{vbag}     M.~Cierniak, T.~Klähn, T.~Fischer and N.~U.~Bastian,
                   Universe {\bf 4}, 30 (2018); 
                   T. Kl\"ahn  and T. Fischer,
                   Astrophys. J. {\bf 810}, 134 (2015).

\bibitem{baym19}   Y.~Song, G.~Baym, T.~Hatsuda and T.~Kojo,
                   Phys.\ Rev.\ D {\bf 100}, 034018 (2019).

\bibitem{deb20}    L.~L.~Lopes, C.~Biesdorf, K.~D.~Marquez and D.~P.~Menezes,
                   Phys.\ Scripta {\bf 96}, 065302 (2021). 
                   L.~L.~Lopes, C.~Biesdorf and D.~P.~Menezes,
                   arXiv:2005.13136 [hep-ph].

\bibitem{oert20}   K.~Otto, M.~Oertel and B.~J.~Schaefer,
                   Eur.\ Phys.\ J.\ ST {\bf 229}, 3629 (2020).

\bibitem{pisa21}   R.~D.~Pisarski,
                   Phys.\ Rev.\ D {\bf 103}, L071504 (2021).

\bibitem{sylhz}    L.~Q.~Su, Y.~Yan, C.~M.~Li, Y.~F.~Huang and H.~Zong,
                   Phys.\ Rev.\ D {\bf 103}, 094037 (2021).


\bibitem{antoniadis}     J.~Antoniadis  et al.,
                         Science {\bf 340}, 6131 (2013).
                         
\bibitem{demo10}        P. B. Demorest, T. Pennucci, S. M. Ransom,
                        M. S. E. Roberts,  and J. W. T.  Hessels,
                        Nature {\bf 467}, 1081 (2010).
                         

\bibitem{gui13}          S. Guillot, M. Servillat, N. A. Webb and R. E. Rutledge,
                         Astrophys. J. {\bf 772}, 7 (2013). 

\bibitem{oz16}           F. Ozel, D. Psaltis, T. Guver, G. Baym, C. Heinke
                         and S. Guillot, Astrophys. J. {\bf 820}, 28 (2016).

\bibitem{freire16}       F.  Ozel and P. Freire,
                         Ann. Rev. Astron. Astrophys. {\bf 54}, 401 (2016).


\bibitem{stein17}        A. W. Steiner, C. O. Heinke, S. Bogdanov,
                         C. Li, W. C. G. Ho, A. Bahramian and S. Han,
                         Mon.\ Not.\ Roy.\ Astron.\ Soc.\  {\bf 476}, 421 (2018).

\bibitem{stein16}        J. N\"attil\"a, A. W. Steiner, J. J. E. Kajava,
                         V. F. Suleimanov and J. Poutanen,
                         Astron. Astrophys. {\bf 591}, A25 (2016). 

\bibitem{miller17}       J. N\"attil\"a, M. C. Miller, A. W. Steiner,
                         J. J. E. Kajava, V. F. Suleimanov and J. Poutanen,
                         Astron. Astrophys. {\bf 608}, A31 (2017).

\bibitem{bogda16}        S. Bogdanov, C. O. Heinke, F. \"Ozel and T. G\"uver,
                         Astrophys.   J. {\bf 831}, 184 (2016). 

\bibitem{ligo17}         B. P. Abbott et al.
                         [LIGO Scientic and Virgo Collaborations],
                         Phys. Rev. Lett. {\bf 119}, 161101 (2017). 

\bibitem{croma}          H. Cromartie et. al.
                         Nature Astronomy {\bf 4}, 72 (2020).

\bibitem{ligo20}         R. Abbott et al.
                         LIGO and Virgo Scientific Collaborations,
                         Astrophys. J. Lett. {\bf 896}, L44 (2020).

\bibitem{ligo20a}        B.~P.~Abbott  et al. 
                         LIGO Scientific and Virgo Collaborations, 
                         Astrophys.\ J.\ Lett.\  {\bf 892}, L3 (2020).
                         
\bibitem{nicer1}         M.C. Miller et al.,
                         Astrophys. J. Lett. {\bf 887}, L24 (2019).

\bibitem{nicer2}         T. E. Riley, A. L. Watts, S. Bogdanov, P. S. Ray,
                         R. M. Ludlam, S. Guillot et al.,
                         Astrophys. J. Lett. {\bf 887}, L21 (2019).

\bibitem{nicer3}         S. Bogdanov, S. Guillot, P. S. Ray, M. T. Wolff,
                         D. Chakrabarty, W. C. G. Ho et al.,
                         Astrophy. J. Lett. {\bf 887}, L25 (2019). 
 
\bibitem{nicer4}         Raaijmakers, G. et al.,
                         Astrophys. J. Lett. {\bf 887}, L22 (2019).

\bibitem{nicer5}         C.D. Capano et al., 
                         Nature Astron.\  {\bf 4}, 625 (2020).
 
\bibitem{nicer6}         T.~E.~Riley {\it et al.},
                         arXiv:2105.06980 [astro-ph.HE].
 
\bibitem{ligo18}         R. Abbott et al.
                         LIGO and Virgo Scientific Collaborations,
                         Phys. Rev. Lett. {\bf 121},  161101 (2018).

\bibitem{annala}         Annala, E., Gorda, T., Kurkela, A. and Vuorinen, A., 
                         Phys. Rev. Lett. {\bf 120}, 172703 (2018).
  
\bibitem{davi}           D. A. Foga\c{c}a, F.S. Navarra,
                         Phys. Lett. \ B {\bf 700}, 236 (2011).            

\bibitem{farhi}          E. Farhi, R.L. Jaffe,
                         Phys. Rev. D {\bf 30}, 2379 (1984).

\bibitem{bla20}          M.~Cierniak and D.~Blaschke,
                         Eur.\ Phys.\ J.\ ST {\bf 229}, 3663 (2020).

\bibitem{zzl}            E.~P.~Zhou, X.~Zhou and A.~Li,
                         Phys.\ Rev.\ D {\bf 97}, 083015 (2018).
            
\bibitem{german}         A.~Parisi, C.~V.~Flores, C.~H.~Lenzi,
                         C.~S.~Chen and G.~Lugones,
                         arXiv:2009.14274.

\bibitem{glend}          N. Glendenning,
                         {\it Compact stars}, (Springer, New York, 2000). 
                         
%
                         
\bibitem{tanja}          T. Hinderer,   arXiv: 0711.2420; 
                         E. E. Flanagan and T. Hinderer,
                         Phys. Rev. D {\bf 77}, 021502 (2008);
                         T. Hinderer, Astrophys. J. {\bf 677}, 1216 (2008).
                         
\bibitem{tanja2}         T. Hinderer, B. D. Lackey, R. N. Lang, J. S. Read,
                         Phys. Rev. D {\bf 81}, 123016 (2010).

                         
\bibitem{sabatucci}      A. Sabatucci
                         {\textit{Tidal deformation of neutron stars}},
                         Thesis, La Sapienza – University of Rome  (2018).

\bibitem{latt10}         S.~Postnikov, M.~Prakash and J.~M.~Lattimer,
                         Phys.\ Rev.\ D {\bf 82}, 024016 (2010).
                         
\bibitem{under20}        A.~Zacchi, arXiv:2007.00423.



\bibitem{laura}          O.~Lourenço, C.~H.~Lenzi, M.~Dutra, E.~J.~Ferrer,
                         V.~de la Incera, L.~Paulucci and J.~E.~Horvath,
                         arXiv:2104.07825. 

\bibitem{wsz}            Q.~Wang, C.~Shi and H.~S.~Zong,
                         Phys.\ Rev.\ D {\bf 100}, 123003 (2019); 
                         Erratum: [Phys.\ Rev.\ D {\bf 100}, 129903 (2019)]. 

\bibitem{zm}             C.~Zhang and R.~B.~Mann,
                         Phys.\ Rev.\ D {\bf 103}, 063018 (2021). 

\bibitem{lz}             C.~M.~Li, S.~Y.~Zuo, Y.~Yan, Y.~P.~Zhao, 
                         F.~Wang, Y.~F.~Huang and H.~S.~Zong,
                         Phys.\ Rev.\ D {\bf 101}, 063023 (2020).  



\bibitem{consta}         M.~Ferreira, R.~Câmara Pereira and C.~Providência,
                         Phys.\ Rev.\ D {\bf 102}, 083030 (2020).


\bibitem{juwu}           M.~Ju, X.~Wu, F.~Ji, J.~Hu and H.~Shen,
                         Phys.\ Rev.\ C {\bf 103}, 025809 (2021). 

\bibitem{theo}           T.~F.~Motta, A.~M.~Kalaitzis, S.~Antic, 
                         P.~A.~M.~Guichon, J.~R.~Stone and A.~W.~Thomas,
                         Astrophys.\ J.\  {\bf 878}, 159 (2019). 

\bibitem{kmt}            S.~Khanmohamadi, H.~R.~Moshfegh and S.~Atashbar Tehrani,
                         Phys.\ Rev.\ D {\bf 101}, 123001 (2020).  

\bibitem{han18}        S.~Han and A.~W.~Steiner,
                       Phys.\ Rev.\ D {\bf 99}, 083014 (2019).


\bibitem{latt18}       T.~Zhao and J.~M.~Lattimer,
                       Phys.\ Rev.\ D {\bf 98},  063020 (2018).

\bibitem{tan}            N.~H.~Tan, D.~T.~Khoa and D.~T.~Loan,
                         Eur.\ Phys.\ J.\ A {\bf 57}, 153 (2021).

\bibitem{latt21}         C.~Drischler, S.~Han, J.~M.~Lattimer, 
                         M.~Prakash, S.~Reddy and T.~Zhao,
                         Phys.\ Rev.\ C {\bf 103}, 045808 (2021). 

\end{thebibliography}
\end{document}